\documentclass[aps,prl,reprint,groupedaddress]{revtex4-1}

\usepackage{graphicx}
\usepackage{dcolumn}
\usepackage{inputenc}
\usepackage{amsmath}

\begin{document}

\title{Object Wave Reconstruction by Phase-Plate Transmission Electron Microscopy}

\author{B. Gamm}
\email{gamm@kit.edu}
\affiliation{Laboratorium f\"ur Elektronenmikroskopie, Karlsruher Institut f\"ur Technologie (KIT), 76131 Karlsruhe, Germany}
\author{M. Dries}
\affiliation{Laboratorium f\"ur Elektronenmikroskopie, Karlsruher Institut f\"ur Technologie (KIT), 76131 Karlsruhe, Germany}
\author{K. Schultheiss}
\affiliation{Laboratorium f\"ur Elektronenmikroskopie, Karlsruher Institut f\"ur Technologie (KIT), 76131 Karlsruhe, Germany}
\author{H. Blank}
\affiliation{Laboratorium f\"ur Elektronenmikroskopie, Karlsruher Institut f\"ur Technologie (KIT), 76131 Karlsruhe, Germany}
\author{A. Rosenauer}
\affiliation{Institut f\"ur Festk\"orperforschung, Universit\"at Bremen, D-28359 Bremen Germany}
\author{R.R. Schr\"oder}
\affiliation{Max-Planck-Institute of Biophysics, Max-von-Laue-Str. 3, D-60438 Frankfurt am Main, Germany}

\author{D. Gerthsen}
\affiliation{Laboratorium f\"ur Elektronenmikroskopie, Karlsruher Institut f\"ur Technologie (KIT), 76131 Karlsruhe, Germany}

\date{\today}

\begin{abstract}
A method is described for the reconstruction of the amplitude and phase of the object exit wave function by phase-plate transmission electron microscopy. The proposed method can be considered as in-line holography and requires three images, taken with different phase shifts between undiffracted and diffracted electrons induced by a suitable phase-shifting device. The proposed method is applicable for arbitrary object exit wave functions and non-linear image formation. Verification of the method is performed for examples of a simulated crystalline object wave function and a wave function acquired with off-axis holography. The impact of noise on the reconstruction of the wave function is investigated.
\end{abstract}

%\keywords{}

\maketitle

\section{Introduction}
Recent advances in the development of physical phase plates have opened new imaging capabilities in transmission electron microscopy (TEM). There are different approaches to realize such phase plates, for example the electrostatic Boersch phase plate \cite{Boersch1947,Schultheiss2006}, drift-tube phase plate \cite{Cambie2007}, or carbon-film-based Zernike or Hilbert phase plates \cite{Danev2001,Danev2004}. Except for the Hilbert-type phase plate, all these phase plates have in common that they aim at shifting the relative phase of the diffracted and undiffracted electrons. In the weak-phase object approximation phase plates provide contrast enhancement at small and intermediate spatial frequencies compared to normal bright-field images. However, in any TEM image only the absolute square of the image wave function is observed with the consequence that all phase information is lost. Moreover, due to imperfect lenses the image wave function is modified by aberrations compared to the object wave function, meaning that information at spatial frequency u of the object is transferred with a different amplitude and different phase into the image. This also occurs in phase-plate imaging, even in combination with an aberration corrector. A constant phase-contrast transfer function of one would result in a constant amplitude contrast transfer function of zero. To retrieve the full phase and amplitude information of the original complex object wave function is therefore a general problem in transmission electron microscopy. Since Gabor's \cite{Gabor1948,Gabor1949} proposal for holography being used in TEM, several methods were proposed and several were realized experimentally. For material science specimens off-axis holography is a well known and quite useful method for amplitude and phase reconstruction \cite{Tonomura1982,Lichte1986}. However, the small area which can be reconstructed is a disadvantage of off-axis holography. This restriction can be alleviated by a Lorentz lens but at the expense of spatial resolution. Another technique relies on through-focus series, where several images at different defoci need to be taken \cite{Coene1996,Thust1996,Koch2008}. This in-line holography method requires exact determination of aberrations and defocus of each image for correct reconstruction. 
The method proposed by Danev and Nagayama \cite{Danev2001a} uses a Zernike- or Hilbert-type phase plate to reconstruct the object wave function for weak-phase objects, which are commonly investigated in biology. Images taken with phase shifts induced by a phase plate we refer to as 'phase-contrast images' in the following. According to Danev and Nagayama, phase retrieval of weak-phase objects requires a phase contrast and conventional TEM image. The image wave function of a weak-phase object is obtained by adding the Fourier-transformed conventional image and complex conjugate of the Fourier-transformed phase-contrast image. The object wave function is obtained by subsequent correction of the image wave function with the wave transfer function.

In this report we present a method for the reconstruction of the complex object wave function of arbitrary objects which do not have to be necessarily weak-phase objects. The method requires the use of a physical phase plate or any other device which is capable of imposing at least two different relative phase shifts between undiffracted and diffracted electrons. Three  images need to be taken at three different arbitrary phase shifts, for example at $-\pi/2,$ 0 and at $\pi/2$. All other parameters like defocus, C$_{S}$-value and exposure time must be kept constant in all three images. The technique allows a full analytical reconstruction of the object exit wave function. The complex wave function can be reconstructed without any restrictions because the reconstruction does not only work for weak-phase, but also for strong-phase and/or strong-amplitude objects. It also does not matter if the images are dominated by linear image formation or if they contain non-linear contributions. The validity of the approach is demonstrated by an experimental wave function, which was obtained by off-axis holography. Moreover, non-linear image formation was taken into account to calculate images of a simulated crystalline object wave function. The images were calculated for different phase shifts of the undiffracted electrons. The reconstruction process based on such three images yields the local amplitude and phase which coincide well with the original wave function.

\section{Experimental Techniques}
To obtain a real test wave function for the reconstruction process, off-axis holography in a transmission electron microscope was applied to reconstruct the local amplitude and phase of a test sample consisting of nanoscaled Pt-particles dispersed on a thin amorphous carbon (a-C) film. For off-axis electron holography, a Philips CM200 FEG/ST was used which is equipped with a Möllenstedt biprism, i.e. a thin wire, in the selected-area aperture holder. The electrostatic potential at the biprism was set to approximately 180 V corresponding to an interference fringe distance of 0.2 nm. Holograms were obtained by inserting the sample into the beam in such a way that it is positioned on one side of the biprism with the specimen edge approximately parallel to the biprism. The object wave is transmitted through the sample whereas the reference wave propagates through vacuum in the hole of the TEM specimen on the other side of the biprism. Kinematical diffraction conditions prevail for the test sample. 
The TEM image simulation routine of the STEMSIM program \cite{stemsim} written in Matlab (The Mathworks Inc.) was extended with the possibility to define an additional phase shift $\phi_{PP}$ between diffracted and undiffracted electrons. It was also complemented by a software package to implement the off-axis holography reconstruction as well as the phase-plate reconstruction. More experimental details and the off-axis holography reconstruction procedure are outlined by Lehmann and Lichte \cite{Lehmann2002}.

\section{Wave function reconstruction by phase-contrast transmission electron microscopy}
The information of the object after transmission through the sample is encoded in the object wave function given by $ \Psi(\vec{r})=\left| \Psi_{0}\right| exp(i \phi_0 (\vec{r}) )- \left|\Psi_{D} (\vec{r}) \right| exp(i \phi_D (\vec{r}) )$ which is composed of the local ampltiudes and phases of the unscattered $\Psi_{0}$ and the diffracted electron wave $\Psi_{D}$. The object wave function is modified by the properties of the imaging lens system which imposes a phase shift $\chi(u)= \pi(Z\lambda u^2+ \frac{1}{2} C_S \lambda^3 u^4 )$ on the object wave. The wave aberration function $\chi$ depends on the spatial frequency u and on the objective lens aberrations such as defocus Z and spherical aberration coefficient C$_ {S}$.
With a phase plate in the back focal plane of the objective lens we can change the phase shift of the undiffracted electrons $\phi_0 (\vec{r})$ at u=0 in the case of a Boersch phase-plate by varying electrode voltage. To consider the effect of an ideal phase-shifting device on image formation, the relative phase shift between diffracted and undiffracted electrons is taken into account by including $\phi_{PP}$ in the wave aberration function.
\begin{equation}\label{eqn1}
\begin{split}
u \neq 0: &\chi(u)= \pi (Z\lambda u^2+1/2 C_S \lambda^3 u^4 ) \\
u=0: &\chi(u)=\phi_{PP}
\end{split}		
\end{equation}
Image formation for arbitrary strong scattering objects is commonly described by the transmission cross-coefficient (TCC) formalism \cite{Ishizuka1980}. The intensity (of reflections) in the Fourier-transformed image at a spatial frequency u is given by
\begin{equation}\label{eqn2}
 I(u)=\int \Psi(u+u' ) \Psi^* (u' )~T(u+u',u' ) du'
\end{equation}
with the wave function in the diffraction plane Ψ(u). The transmission cross-coefficient T(u+u',u') given by 
\begin{equation}\label{eqn3}
T(u+u',u' )=T(u'',u' )=E(u'',u' ) \cdot exp[i \cdot (\chi(u'' )-\chi(u' ) )]
\end{equation}
which takes into account the phase shift induced by the objective lens aberrations $\chi(u)$. The effects of partial spatial and temporal coherence are described by envelope functions contained in E. More details on the explicit form can be found in \cite{Ishizuka1980,Chang2005}.

The expression given by Eq. (\ref{eqn2}) can be separated into the linear contributions (i.e. the interference between undiffracted and diffracted electrons u'=0, u'=-u) and non-linear contributions (i.e. the interference between diffracted electrons ($u'\neq'0$, $u'\neq-u$).
\begin{equation}\label{eqn4}
\begin{split}
I(u)&=\Psi(u) \Psi^* (0)~T(u,0)+\Psi^* (-u) \Psi(0) T^* (-u,0)  \\&+ \int_{u'≠-u,u'≠0} \Psi(u+u') \Psi^* (u')~T(u+u',u')du'+I' (u)
\end{split}
\end{equation}
I'(u) represents the sum of all contributions to the image intensity which are incoherent with respect to the unscattered electron and are not considered by the integral in Eq. (\ref{eqn4}). We will comment on I'(u) below. 
The integral in Eq. (\ref{eqn4}) can be neglected for weak-phase objects which yields only negligible contributions to the image intensity. Non-linear contributions in images of strong objects prevent the application of the method presented by Danev and Nagayama \cite{Danev2001a}  for wave function reconstruction which is extended in this work for arbitrary objects. Our method outlined in the following subchapters relies on the obvious fact that a phase shift imposed on the phase of the unscattered electrons by a phase plate changes only the intensities of the linear contributions leaving the non-linear ones unaffected.
I'(u) represents contributions to the image intensity which are incoherent with respect to the unscattered electrons. This applies to all inelastically scattered electrons because a (stationary) phase relationship does not exist between unscattered and inelastically scattered electrons. We note that inelastically scattered electrons may produce interference patterns as shown by energy-filtered off-axis holography \cite{Potapov2006}. However, these interference patterns cannot be exploited for wave function reconstruction by phase-plate imaging because zero-loss and energy-loss interference patterns add up incoherently. 

\begin{figure}
\includegraphics[width=0.45\textwidth]{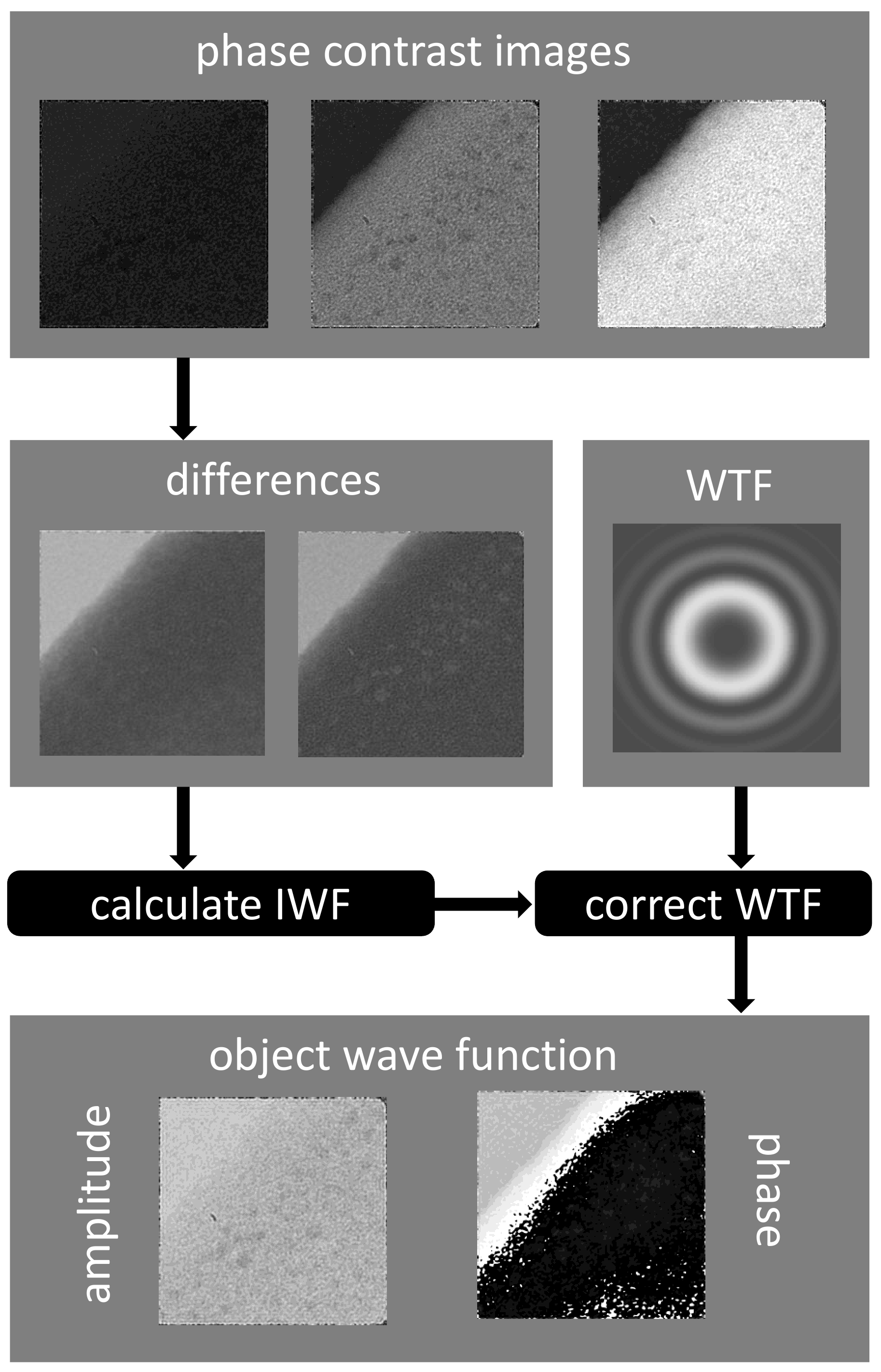}
\caption{\label{fig:fig1}Schematic illustration of the required reconstruction steps for the presented method. Input parameters are three aligned images and the complex wave transfer function. Image wave function (IWF) is calculated and corrected by the WTF. Amplitude and phase of the exit object wave function are the result.}
\end{figure}

In the following two sections we describe reconstruction formalisms for object wave function in real and reciprocal space. To give an overview over the steps needed for reconstruction an illustration is given in Figure \ref{fig:fig1}. Starting with the three acquired images with different phase shifts applied by the phase-plate, the two differences are calculated after an alignment procedure to correct for specimen drift. From these differences the image wave function (IWF) is calculated and corrected with the wave transfer function. The result is the complex form of the object exit wave function.
%%%figure 1
%% end figure1

\subsection{Reconstruction in real space}
The reconstruction of the image wave function requires three images taken with the same imaging parameters, but different phase shifts of the undiffracted beam. If not only the image wave function but also the object wave function needs to be retrieved, the aberrations of the imaging system need to be known. A wave transfer function correction can be carried out as described e.g. by Lehmann and Lichte \cite{Lehmann2002}. If a C$_S$-Corrector is available C$_S$ and defocus should both be set to the smallest possible value for all three images, which might make a wave transfer correction unnecessary, depending on the slope of the wave aberration function in the range of spatial frequencies used for reconstruction. The image intensity is determined by the sum of linear and non linear contributions. Neglecting aberrations for the moment, as they can be corrected after the reconstruction, we can describe the image intensity in real space by the following equation
\begin{equation}\label{eqn5}
I(\vec{r})=\left| \Psi_0 - \Psi_D (\vec{r}) \right|^2
\end{equation}

$\Psi_0$ contains the phase and amplitude of the undiffracted electrons and $\Psi_D (\vec{r})$ contains the contributions of all diffracted electrons at location $\vec{r}$. In the three required images, $\phi_ {PP}$ for example may be set to 0, $\pi/2$ and $- \pi/2$. Although the phase shifts can be arbitrary, these particular phase shifts were chosen to simplify the following equations. Moreover, the signal-to-noise ratio will be better if the difference between minimum and maximum phase shift is $\pi$ because we see contrast reversal between the two extremes. Note, that the contrast enhancing properties of phase-contrast imaging with a physical phase plate, rely totally on weak-phase objects and the associated approximation. For strong-phase objects, the contrast can decrease for certain $\phi_{PP}$. It is important for the image acquisition that the exposure time is identical for all three images, and the requirements for experimental conditions apply as discussed in chapter 5. In the next step the difference between each of the two phase-contrast images and the "conventional" image is calculated. The phase of the undiffracted part of the wave $\Psi_0$ is set to zero, because only phase differences are of interest, while absolute phases are not observable. The result is the difference between the linear interference terms only, as all terms which are independent of the phase shift of the undiffracted beam are eliminated. The following equations describe these intensity differences 
\begin{equation}\label{eqn6}
\begin{split}
\Psi(\vec{r})=&\left| \Psi_0 \right| - \left|\Psi_D (\vec{r}) \right| e^{i \phi(\vec{r})} \\
\Psi_{(\pm \pi/2)} (\vec{r}) =&\left| \Psi_0 \right| e^{(\pm i \pi/2)} - \left| \Psi_D (\vec{r}) \right| e^{i \phi(\vec{r})} \\
D_{(0,\pm \pi /2)} (\vec{r}) &=\Psi (\vec{r}) \Psi^* (\vec{r}) - \Psi_{(\pm \pi/2)} (\vec{r}) \Psi_{(\pm \pi/2)}^* (\vec{r}) \\
=& \left| \Psi_0 \right| \left|\Psi_D (\vec{r}) \right| (e^{i \phi(\vec{r})}  (1\pm i)+e^{(-i \phi(\vec{r}))} (1\mp i))
\end{split}
\end{equation}           
with the amplitude of the undiffracted wave $\left|\Psi_0 \right|$ and the amplitudes and relative phases of the diffracted part of the wave function $ \left| \Psi_D (\vec{r}) \right|$ and $\phi(\vec{r})$. It should be noted that $\phi(\vec{r})$ denotes a phase linked with the exit wave function, whereas the phase shift induced by the physical phase-plate is always denoted as $\phi_{PP}$. Since the phase shift $\phi_{PP}$ was selected to be $\pi/2$ and $-\pi/2$, its effect is contained in the signs and complex number i in the brackets of the equations. The equations contain linear image contributions because all non-linear contributions as well as incoherent (with respect to the undiffracted electrons) contributions are not affected by the phase shift that was imposed on the undiffracted wave. Constant terms which are independent of the phase shift of the undiffracted electrons are eliminated. The system of equations can be solved, if a third equation is available which is linearly independent and contains separately $\left| \Psi_0 \right|$  and $\left| \Psi_D (\vec{r}) \right|$ . So far we can solve the system of equations: 
\begin{equation}\label{eqn7}
\begin{split}
\frac{1}{\gamma}  (D_{(0,-\pi/2)}+iD_{(0,+\pi/2)} )=\\e^{i \phi} (1-i+i-1)+e^{-i \phi} (1+i+i+1)=&2e^{-i \phi} (1+i)\\
\frac{1}{\gamma}  (iD_{(0,-\pi/2)}+D_{(0,+\pi/2)} )=\\e^{i\phi} (i+1+1+i)+e^{-i\phi} (i-1+1-i)=&2e^{i\phi} (1+i) \\
with~ \gamma=\left|\Psi_0 \right|\left|\Psi_D \right|
\end{split}
\end{equation}

for $\phi (\vec{r})$ and for $\left| \Psi_D (\vec{r}) \right|$ dependent on $\left| \Psi_0 \right|$, which are given as:

\begin{equation}\label{eqn8}
\begin{split}
e^{-i \phi}&=\sqrt{\frac{D_{(0,-\pi/2)}+ i D_{(0,+\pi/2)}}{iD_{(0,-\pi/2)}+{D_{(0,+\pi/2)}}}} \\
\phi&= \frac{1}{2} i ln(\frac{D_{(0,-\pi/2)}+iD_{(0,+\pi/2)}}{iD_{(0,-\pi/2)}+D_{(0,+\pi/2)}}) \\&
\left| \Psi_D \right| = \frac{e^{-iφ}}{2(1+i)\left|\Psi_0 \right|} (iD_{(0,-\pi/2)}+D_{(0,+\pi/2)})
\end{split}
\end{equation}

The phase $\phi(\vec{r})$ of the diffracted electrons is independent of $\left|\Psi_0 \right|$, but it cannot be translated directly into the phase of the exit wave function, because it has the form
\begin{equation}\label{eqn9}
\Psi(\vec{r}) = \left|\Psi_0 \right|-\left|\psi_D (\vec{r}) \right| e^{i\phi(\vec{r})}
\end{equation}

Different approaches are possible to obtain a third equation for $\left|\Psi_0 \right|$ , but they all require some knowledge of what is seen in the image. In the following we present a method which requires a small vacuum region in the image where the square of the exit wave function is unity. We can then normalize the reconstructed image wave function for an illumination with a plane wave with unity amplitude by setting $\left|\Psi_0 \right|=1$ and dividing the whole image set by the vacuum intensity.
If a vacuum region is not present in the image due to experimental restrictions, it may be possible to find other boundary conditions for the system of equations. For example if the amplitude in a certain area (i.e. carbon film or another homogenous area) is known. It should be noted that the vacuum could be captured by a fourth image away from the region of interest, but the dynamic range and/or bit-depth (12-bit,14-bit etc.) of the CCD-camera might prevent such an image at exactly the same conditions due to over-exposure.
If the reconstruction is carried out in real space the full spectrum of spatial frequencies contained in the image is used for reconstruction. For some applications, especially for images taken without a C$_S$-corrector, it might be useful to reconstruct only up to a certain spatial frequency u. It should be noted, that the same reconstruction method can be carried out in Fourier space, with the exception of u=0, for which $\left|\Psi_0 \right|$ has to be inserted. During the reconstruction process the spatial frequency interval can be easily limited by use of an aperture function.
After successful reconstruction of the image wave function (IWF), the object wave function can be obtained by taking the wave transfer function (WTF) into account in analogy to off-axis holography \cite{Lehmann2002} where the Fourier-transformed image wave function is divided by the wave transfer function in Fourier space. Since aberrations should be constant during the acquisition of the three images, only one set of parameters needs to be determined. It is likely that this can be done with improved precision from three images, if of course the phase shift is taken into account. If an aberration-corrected microscope is used, it is possible to correct all aberrations as precisely as possible. Since phase contrast is produced by the phase plates, it is not necessary to have other than residual aberrations for phase contrast transfer \cite{Gamm2008}. In this case no WTF-correction is needed if the wave aberration function is flat over the reconstructed part of the Fourier space. 

\subsection{Reconstruction in Fourier space}
Based on the same principle the entire reconstruction of the object wave function can also be carried out in Fourier space. Again three images of the specimen, taken at the same imaging parameters but at different phase shifts of the undiffracted electrons are required. Although these phase shifts can be chosen arbitrarily, we once again choose 0, $\pi /2$ and $-\pi /2$ to simplify the following equations. Unlike the real-space method presented in the previous chapter, the reconstruction in Fourier space does not rely on a vacuum region in the images of the specimen themselves. It is sufficient to acquire a vacuum image with the same imaging parameters as for the images required for the reconstruction. This vacuum image can be used to normalize the three images of the specimen with respect to the illumination by a plane wave of amplitude 'one'. 
Subsequently the differences $D_{(0,\pm \pi/2)} (u)$  between the Fourier transformed 'conventional' image and each of the Fourier-transformed phase contrast images are calculated. Using Eq. (\ref{eqn3}) and (\ref{eqn4}) these differences can be evaluated for all $u\neq0$:
\begin{equation}
\begin{split}
D_{(0,\pm \pi/2)} (u)=&I_0 (u)-I_{(\pm \pi/2)} (u) \\ =& \Psi(u) \Psi^* (0) E(u,0) exp(-i \chi(u) )(1\pm i)\\+&\Psi(0) \Psi^* (-u) E(u,0)exp(i\chi(u) )(1\mp i)\\,& u\neq0
\end{split}
\end{equation}

with the envelope damping function $E(u'',u')$ and the phase distortion function $\chi (u)$. Again all non-linear contributions vanish as they are independent of the phase shift of the undiffracted beam. Since $D_{(0,\pm \pi/2)} (u)$ are complex-valued the above system of equations can be solved for the complex quantity
\begin{equation}\label{eqn11}
\begin{split}
\Psi(u) \Psi^* (0) E(u,0) exp(i \chi(u) )&= \frac{D_{(0,+\pi/2)}+i D_{(0,-\pi/2)}}{2(1+i)} \\&, u \neq0 
\end{split}
\end{equation}
Correction for aberrations and negligence of the absolute phase of the undiffracted beam (setting the phase of the undiffracted beam equal to zero) leads to $\Psi(u) \cdot\left| \Psi(0) \right|$ for all $u \neq 0$.
The modulus of the undiffracted beam $\left| \Psi(0) \right|$ arises from the integration of the absolute square of $\Psi(u) \cdot \left| \Psi(0)\right|$ over the whole Fourier space except for u=0:
\begin{equation}\label{eqn12}
\sum=\left| \Psi (0) \right|^2 \cdot \int_{u' \neq 0} \left| \Psi (u') \right|^2 du'                                   
\end{equation}
and the intensity I(0) of the Fourier transformed images at u=0. I(0) can be evaluated using equation Eq. (\ref{eqn4}):
\begin{equation}\label{eqn13}
I(0)= \int \left| \Psi (u') \right|^2 du' = \left| \Psi(0) \right|^2 + \int_{u'\neq0} \left|\Psi(u') \right|^2 du'.
\end{equation}
The above system of equations can be solved for $\left| \Psi(0)\right|$: 
\begin{equation}\label{eqn14}
\left|\Psi(0) \right| =\sqrt{\frac{1}{2} (I(0)\pm \sqrt{I(0)^2-4 \sum})}                                   
\end{equation}
Since Eq. (\ref{eqn12}) and (\ref{eqn13}) are quadratic equations, two solutions exist, of which we found the negative signed to be the appropriate one.
Using the quantity $\Psi(u) \cdot \left|\Psi(0)\right|$, obtained by Eq. (\ref{eqn11}), this finally leads to the Fourier transformed object wave function Ψ(u) for all u including u=0. The object wave function in real space $\Psi(r)$ is obtained by inverse Fourier transformation.
Although the reconstruction of the object wave function in Fourier space is based on the same principle as the reconstruction in real space, it is quite sensitive towards the damping envelope functions at higher spatial frequencies, due to the aberration correction during the reconstruction process \cite{Downing2008}. An aperture function should be set to prevent a strong influence of the damping envelope at higher spatial frequencies. It will reduce the spatial resolution of the reconstruction slightly, because it has to be slightly smaller than the corresponding aperture of the spatial resolution of the original images.

\section{Verification of the reconstruction technique}
To obtain a real test wave function for the reconstruction procedure off-axis holography was used. The reconstruction was carried out taking into account spatial frequencies up to $3~nm^{-1}$. The test sample consisted of nanoscaled  Pt-particles on an amorphous (a-)C film. No reflections were visible in the sidebands of the holograms. The Pt-particles have a diameter of $1.5-2~nm$ which corresponds to a phase shift of $\approx 0.12 \pi$. The a-C carrier film induces a phase shift of about $2/3 \pi$ with respect to the vacuum wave. The amplitude is 1 in the vacuum and decreases to about 0.85 for the carbon film. Based on these observations the  Pt-particles can be considered as weak-amplitude strong-phase objects. Strong reflections are not observed in the diffraction pattern, and the intensity of undiffracted electrons is much larger than any diffracted intensity. Linear image formation therefore applies to this particular test object. 
%%%figure 2
\begin{figure*}
\includegraphics[width=0.8\textwidth]{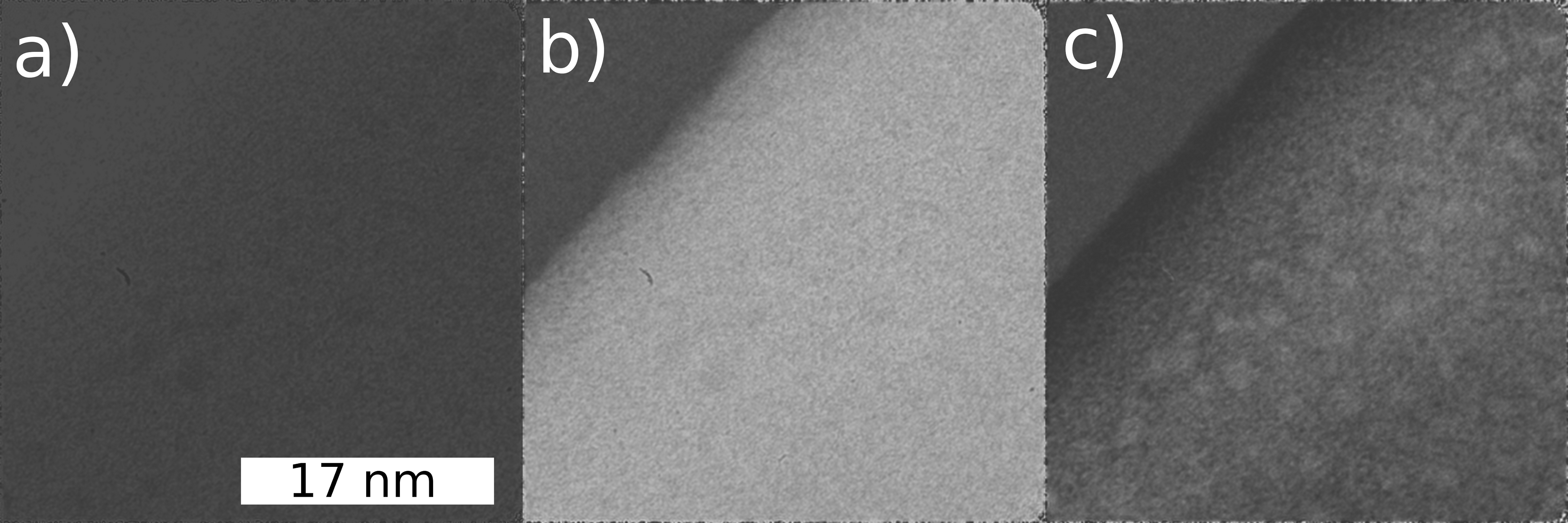}
\caption{\label{fig:fig2}Simulated images calculated on the basis of an object wave function of platinum particles on a thin amorphous carbon film acquired with off-axis electron holography. a) conventional image $\phi_{PP}=0$, b) $\phi_{PP}=\pi/2$, c) $\phi_{PP}=-\pi/2$.  (C$_S$ = 1 mm, defocus Z =  -150 nm, 200 keV)}
\end{figure*}
%% end figure2
The object wave function obtained by off-axis holography will be denoted in the following as original wave function. Figure \ref{fig:fig2} shows three images which are calculated on the basis of the amplitude- and phase-distribution of the test object. The images were simulated assuming linear image formation and different phase shifts $\phi_{PP}= 0$ (Fig. \ref{fig:fig2} a)), $\phi_{PP}=\pi/2$ (Fig. \ref{fig:fig2} b)) and $\phi_{PP}=-\pi/2$ (Fig. \ref{fig:fig2} c)) in the wave aberration function. An ideal phase shifting device without a blocking structure was assumed. The images are calculated for a 200 keV transmission electron microscope with $Z = -150~nm$ and $C_{S} = 1~mm$.  For better comparison of the contrast generated by the phase plate, the contrast in all three images is displayed with the same minimum and maximum gray scale value as well as linear scaling between them. This results in a dark image for the case of $\phi_{PP}=0$ (Fig. \ref{fig:fig2} a)) because the (a-)C film generates only weak contrast at such a comparatively low defocus value.  On the other hand, if the phase-plate phase shift is $\pm \pi/2$ the film generates strong bright contrast (Fig. \ref{fig:fig2} b)) or moderately bright contrast (Fig. \ref{fig:fig2} c)). The vacuum region in the upper left corner shows a constant grey value in all three images, and can be clearly distinguished from the film in Fig. \ref{fig:fig2} b), c). In Fig.\ref{fig:fig2} a) the Pt-particles show dark contrast vs. the film, which is hardly visible due to the chosen contrast scaling. They show a slight dark contrast with respect to the bright (a-)C  film in Fig. \ref{fig:fig2} b) and distinctly bright contrast in comparison to the  a darker (a-)C film in Fig. \ref{fig:fig2} c). Due to the weak-amplitude character of the Pt-particles, the contrast between the film and the particles depends on the phase shift induced by the specimen but also on the phase-plate phase shift. For arbitrary phase objects like this specimen, the optimal contrast can be found anywhere between 0 and $\pm \pi/2$, in contrast to weak-phase objects, where  the optimum contrast is always observed at phase shifts of $\pm \pi/2$. To be more specific, the contrast of the Pt-particles with respect to the (a-)C film changes drastically, because the film only shows little amplitude contrast, but induces a strong phase shift on the electrons, whereas the phase shift of the Pt-particles themselves is only weak compared to that of the film. This is more obvious from the original phase distribution of this particular wave function which is shown in Fig. \ref{fig:fig3} a).
%%%figure 3
\begin{figure*}
\includegraphics[width=0.8\textwidth]{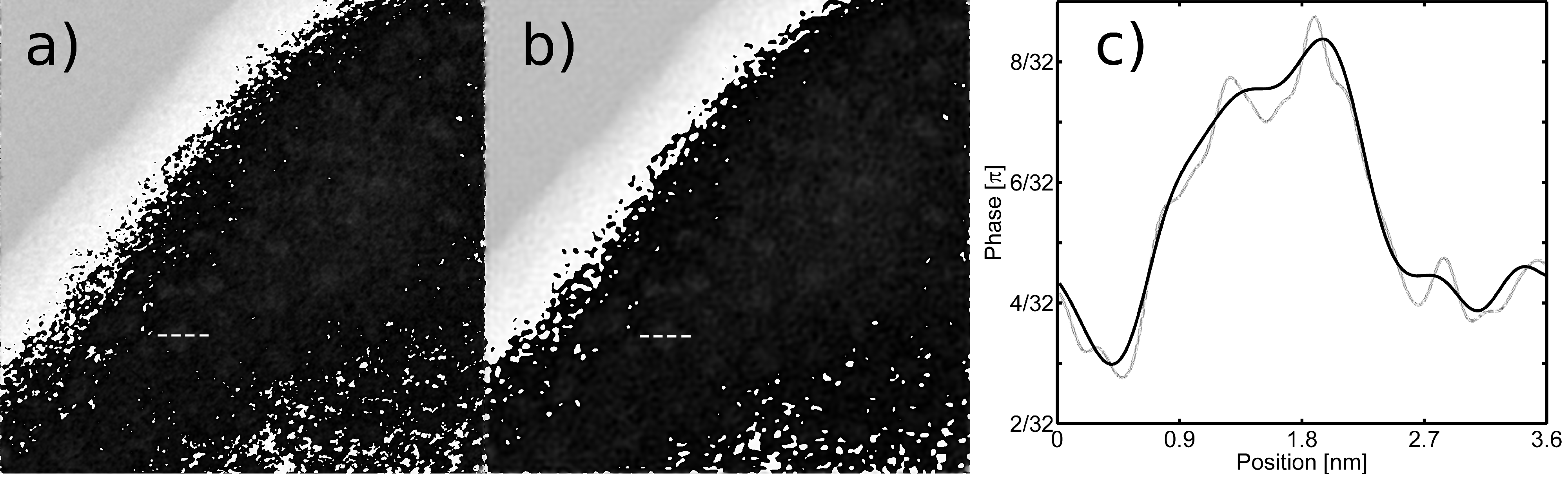}
\caption{\label{fig:fig3}: a) Phase of the original wave function acquired with off-axis electron holography scale from$ -\pi$ (white) to $+ \pi$ (black). b) reconstructed phase from the three images simulated in figure \ref{fig:fig2}). c) Line scans across a particle, taken from  a) (gray line) and taken from b) (black line). The position is marked in a) and b) by the white dashed line.}
\end{figure*}
%% end figure3

Using the reconstruction method outlined in the previous chapter, the phase- and amplitude-distributions can be reconstructed and compared with the original wave function. Fig. \ref{fig:fig3} shows the original (Fig. \ref{fig:fig3} a)) and reconstructed phase distribution (Fig. \ref{fig:fig3} b)). With the exception of a constant phase shift, which results from setting the phase of the undiffracted electrons to zero during the reconstruction, the phase distribution is well reconstructed from the three images. Line scans across a Pt  particle are shown in Fig. \ref{fig:fig3} c) for the original wave function in gray and for the reconstructed wave function in black. Apart from the slightly lower resolution due to an aperture function, the phases are in very good agreement.

For the verification of the reconstruction from images containing non-linear image contributions, another test object wave function is chosen. The object wave function for single crystalline silicon along the [100]-zone axis with a thickness of 10 unit cells corresponding to 5.431 nm was calculated using the STEMsim program \cite{stemsim}. Vacancies were introduced to generate differences in the local phase shift. The vacancies in the atom columns were introduced by removing silicon atoms in the phase gratings for the multislice algorithm. To simulate an occupancy of 0.5 in a particular column, phase gratings of a perfect and defective Si(100) plane were used alternately. The super-cell contains in addition a vacuum region which is required for the reconstruction method in real space. Figure 4 shows images for different phase shifts of a) $\phi_{PP}=0$, b)  $\phi_{PP}=\pi/2$ and c) $\phi_{PP}=-\pi/2$, calculated using an equally spaced numerical integration algorithm tailored for phase-contrast imaging. A 200 keV transmission electron microscope with $C_{S} = 0.5~mm$ and $Z = 5~mm$ was assumed. Based on the three images the image wave function was reconstructed. The object wave function was retrieved by WTF-correction. Figure \ref{fig:fig5} shows the original (Fig. \ref{fig:fig5} a)) and reconstructed phase-distribution (Fig. \ref{fig:fig5} b)), as well as line scans (Fig. \ref{fig:fig5} c)) across an atom column containing vacancies and two adjacent fully occupied atom columns. The atom columns are located at regions with a large phase shift displayed 'in white'. Apart from little noise which was simulated into the images (see next chapter), the reconstructed phase is in good agreement with the original phase used for the image simulation. It should also be noted, that due to the strong amplitude contrast in these images, the contrast between fully occupied atom columns and those with vacancies is barely visible by eye, but is reconstructed correctly, as shown in the line scans in Fig. \ref{fig:fig5} c).
%%%figure 4
\begin{figure}
\includegraphics[width=0.48\textwidth]{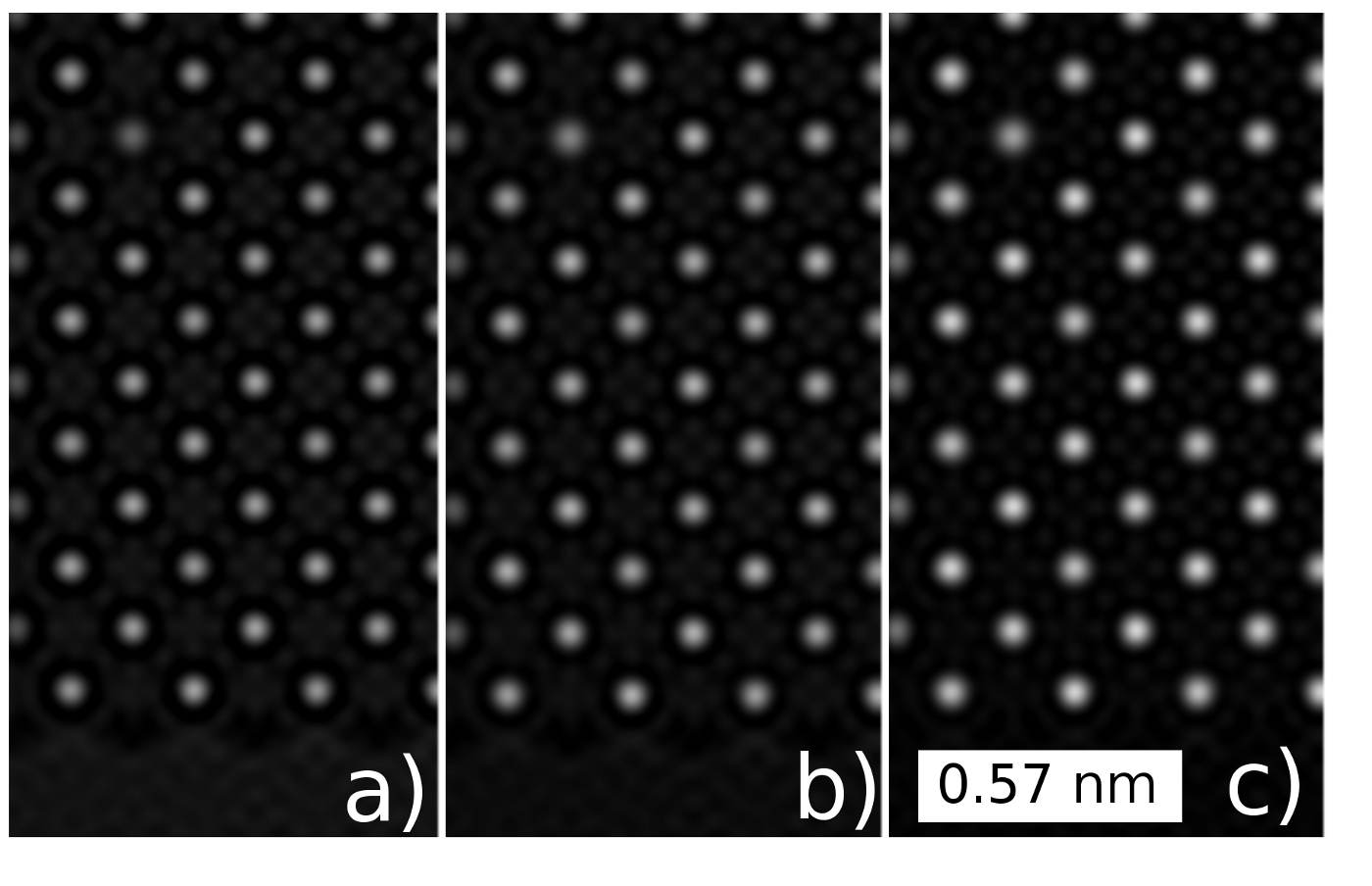}
\caption{\label{fig:fig4}Simulated high-resolution images of silicon crystal along the [100]-zone axis containing vacancies (upper left corner) calculated with TCC formalism for a) conventional image $\phi_{PP}=0$, b) $\phi_{PP}=\pi/2$ , c) image with $\phi_{PP}=-\pi/2$ assuming a 200keV electron microscope with $C_S = 0.5~mm$ and $Z =  5 nm$}
\end{figure}
%% end figure4
%%%figure 5
\begin{figure}
\includegraphics[width=0.48\textwidth]{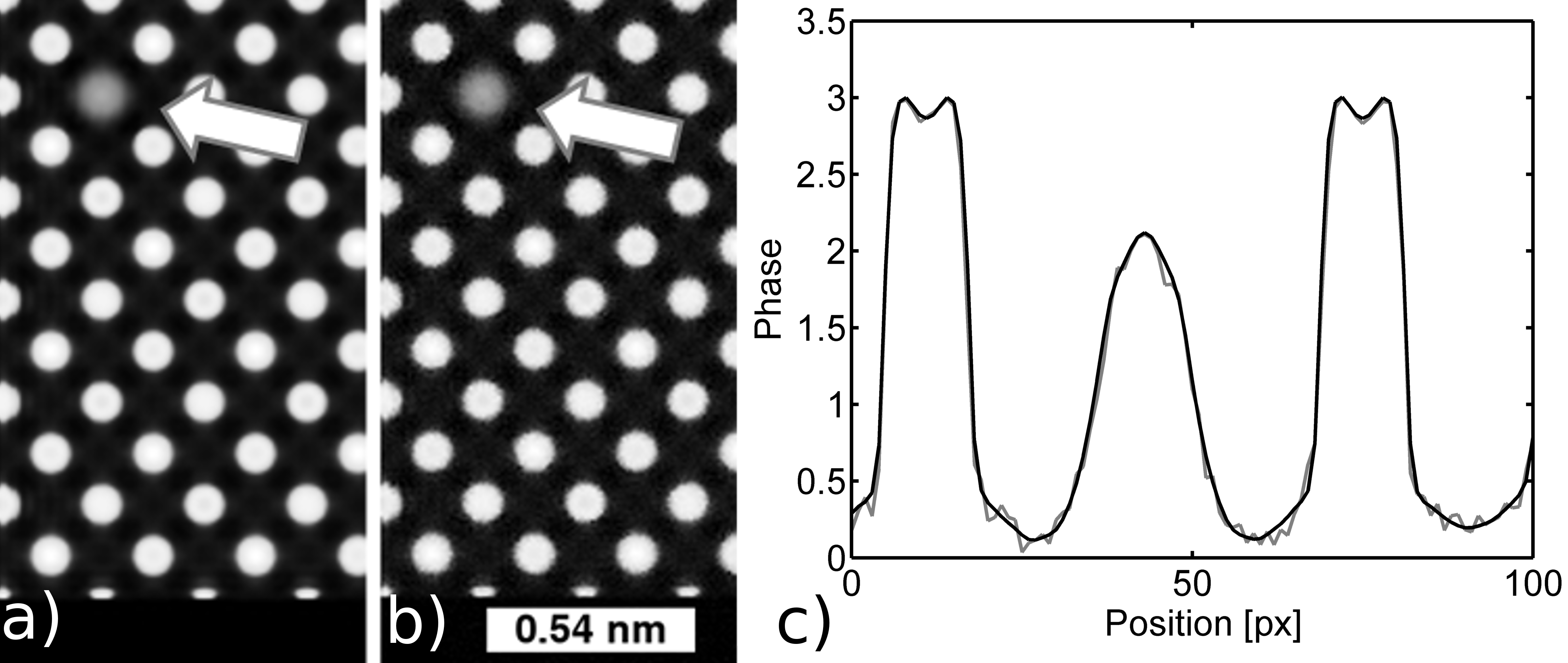}
\caption{\label{fig:fig5}: a) Phase of the original wave function of the silicon sample used to simulate the three images in Figure \ref{fig:fig4}, b) from the three images reconstructed phase of the wave function. White arrows point at the vacancy position. c) Line scans across two atom positions. Black taken from a) across the vacancy position and two adjacent atom positions, and gray taken from b) at the same location. Apart from little noise the line scans are in very good agreement, and show that the reconstruction works even if non-linear contributions are present in the simulated images.}
\end{figure}
%% end figure5

\section{Limitations}
\begin{figure}
\includegraphics[width=0.48\textwidth]{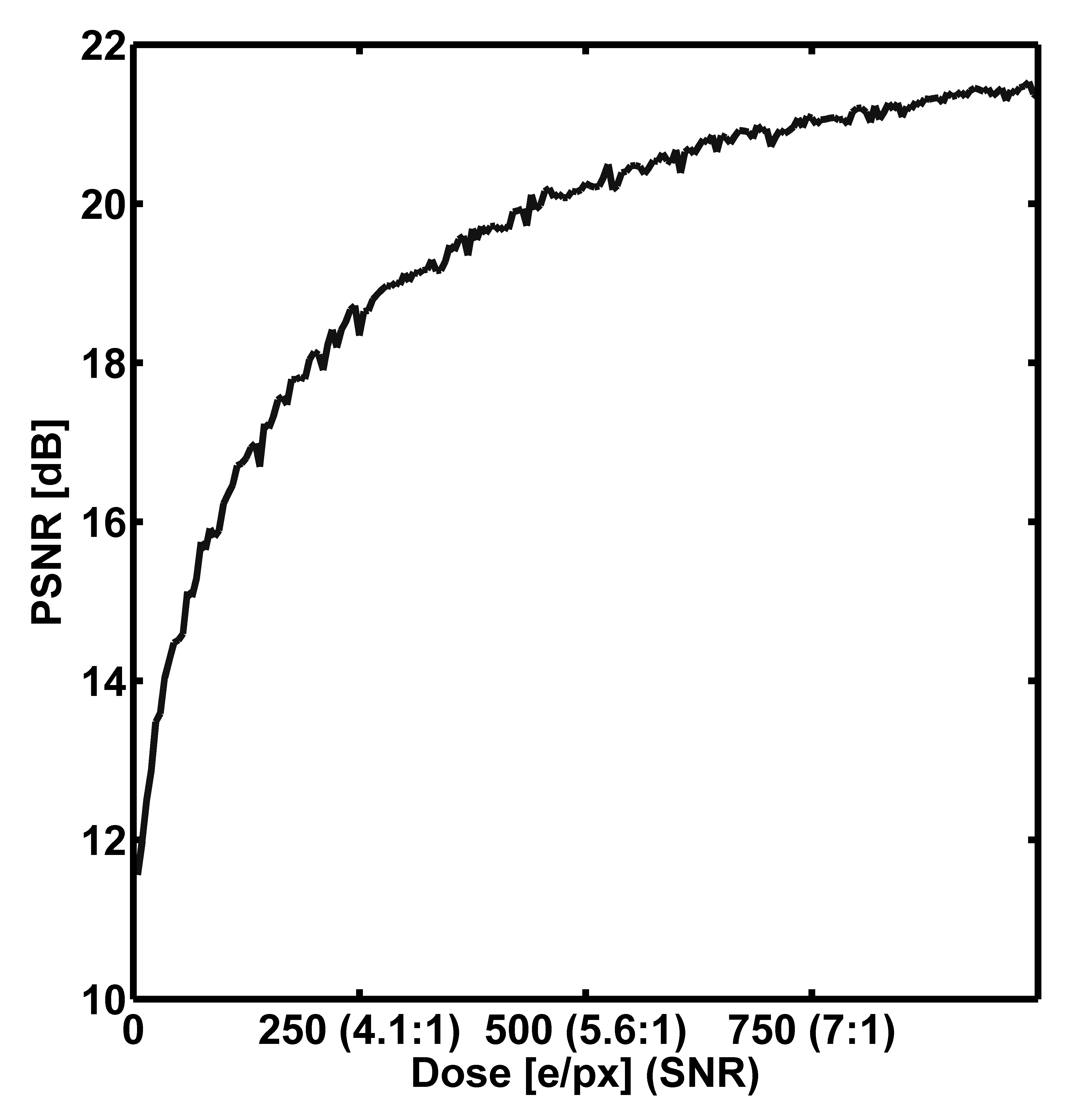}
\caption{\label{fig:fig6}: Peak Signal-to-Noise Ratio (PSNR) calculated from reconstructed phase and original phase of a crystalline silicon sample as a function of the different electron doses per pixel. At low doses, significantly lower than 500 e/px, the phase signal degrades. Signal-to-Noise for 250 e/px, 500 e/px and 750 e/px for the Si crystal specimen are given in the brackets. PSNR values above 20dB are acceptable for noisy images.}
\end{figure}

\begin{figure}
\includegraphics[width=0.35\textwidth]{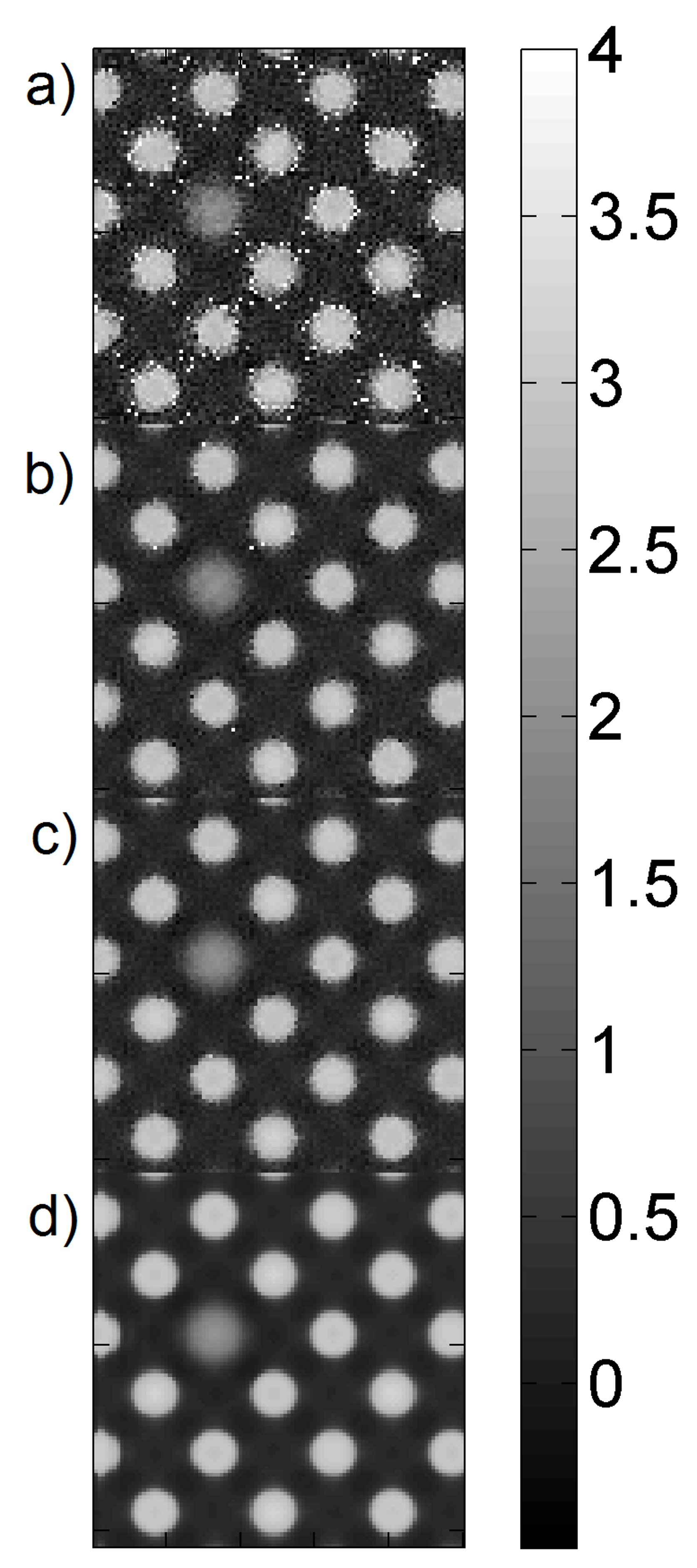}
\caption{\label{fig:fig7}: Reconstructed phase with different electron doses per pixel. a) 250 e/px b) 1250 e/px c) 2500 e/px d) original phase of the wave function. The comparison shows the degradation of the phase signal. As the peak signal-to-noise ratio calculations suggest, noise does not inhibit the possibility to reconstruct the wave function with the proposed method more than is expected.}
\end{figure}

Due to the complexity of the equations used for reconstruction, one might expect that noise completely prevents the reconstruction of the object exit wave function. Therefore a series of image simulations was carried out to simulate the effect of different noise levels. The noise was simulated as Poisson noise for different average electron doses per image pixel. The examination shows that only at lower electron doses below 500 electrons per pixel, a significant effect is observable. However, even at 250 electrons per pixel the phase distribution in a crystalline silicon specimen can be reconstructed properly. Fig. \ref{fig:fig6} shows calculated peak signal-to-noise ratios (PSNR) from the original phase and reconstructed phase for different electron doses. The plot shows a drop in the PSNR at lower electron doses, approximately starting at 500 e/px. At this dose the phase signal starts to degrade, but is still in good agreement with the original phase down to noise levels where the SNR of the images taken is too small at all. While 20dB PSNR are considered acceptable for noisy images, values below this mark show typical artefacts at high noise levels, which cause local deviations from the original distribution. Fig. \ref{fig:fig7} shows a series of reconstructed phases for Si in [100]-zone axis  for different doses and the original phase for comparison. The top image with an electron dose as low as 250 e/px still shows a phase distribution which is in good agreement with the original phase, apart from being much noisier. 250 electrons per pixel approximately translate to a SNR of 4 for the object wave function under investigation. To achieve a high SNR coherence is important. For strong-amplitude specimens, like almost all crystalline objects in zone-axis orientation, the dominant part in the experimental images is likely to be  $\left|\Psi_0 \right|$  and $\left|\Psi_D (\vec{r}) \right|$  and only little contrast remains for the phase part. It can be concluded that a high coherence should reduce the noise in the phase contrast part, which is necessary because it is the only information used for reconstruction. 
%%%figure 6
%% end figure6
Non-linear terms add to this problem, therefore it is also important that the dynamic range and number of grey levels of the camera (i.e. bit depth) is large enough for the specimen under investigation. For weak-phase objects this is not expected to be a problem but this behaviour was often observed for object wave functions with a strong amplitude contribution if an 8-bit camera was assumed in the simulations with a limited number of discrete counts. Of course this depends significantly on aberrations and the spatial-frequency spectrum of the object wave function. Under these conditions a correct reconstruction is impossible and can easily be identified by a very noisy reconstructed phase of the wave function. Coherence and SNR is expected to be one of the most important experimental problems of this method. Very small illumination angles for high spatial coherence and an undiffracted beam with a small diameter requires a field-emission source electron microscope. A high mechanical and electrical phase-plate stability and specimen stability are needed because long exposure times are required for a high SNR. 
%%%figure 7
%% end figure7
Another limitation arises due to the fact that real physical phase plates have different characteristics than assumed here for the reconstruction method. Depending on the type of phase plate, they introduce different perturbations in the back focal plane of the objective lens. In the case of a Boersch phase-plate, the electrode ring blocks a certain range of spatial frequencies. The electrode structure also induces a cut-on frequency at which diffracted electrons start to be phase shifted with respect to the undiffracted electrons. Other phase-plate types may have other characteristics depending on how they shift the phase of selected electrons. 
Simulations were performed with different phase-plate geometries and different shapes of electrostatic potentials in the back focal plane for different object wave functions. Their effect was taken into account by a special shape of the phase-plate function which is a complex function of u: $P(\vec{u})=A(\vec{u})exp(i \phi_{PP} (\vec{u}))$. It was found that the accuracy of the reconstruction depends significantly on how well the perturbation is confined to small spatial frequencies u (near the undiffracted electrons) compared to the relevant information in Fourier space. The reconstruction fails, if a significant fraction of electrons is diffracted into u values which are strongly perturbed by the physical phase plate itself. On the other hand the reconstruction works perfectly, if the diffracted information is contained mainly at spatial frequencies which are not affected by the physical phase plate. A good example for this behavior is a crystalline silicon specimen, where even with vacancies most of the information is contained in relatively large spatial frequencies which are not influenced by a perturbation in the vicinity of the undiffracted beam. This demonstrates that the influence of real physical phase-plates can hardly be generalized. It always depends on the type of phase-plate and the size of the feature of interest.
Electrostatic phase plates offer an significant advantage compared to carbon film phase plates if objects are studied which are not weak-phase objects. Variable phase shifts can be imposed, which is necessary for object wave reconstruction suggested in this work. For arbitrary wave functions, the reconstruction of the amplitude and phase requires three images with different phase shifts. Only two images can be obtained with one single carbon film phase plate, which is not sufficient for a proper reconstruction as demonstrated in chapter 3. It can also be understood from writing the complex function in sine and cosine functions which are not injective functions, and therefore the phase cannot be reconstructed uniquely from only two images. It is often discussed to what extent phase-plates increase information that can be obtained from a single image, mainly by applying the weak-phase-object approximation. This approach is commonly used due to its accessibility and experimental importance to justify experimental efforts made in phase-plate development. Recently, higher order series-expansion has been applied to investigate strong-phase-objects and the increase in contrast of the obtained phase contrast images \cite{Danev2009}. Carbon-film phase-plates with different thicknesses are suggested. However, for arbitrary wave functions, which are of concern to material science, one will immediately conclude, that using a phase plate with a fixed phase shift, whatever phase shift that would be in particular, will result in strong phase contrast or none at all - very much depending on the object wave function itself.

\section{Summary}
Phase-shifting devices with the capability of imposing variable phase shifts can be used to reconstruct the amplitude and phase of wave functions from arbitrary objects on the basis of only three different images. Our method relies on the difference between two images acquired under the same experimental conditions apart from the phase shift. This effectively eliminates all contributions due to non-linear image formation which cannot be avoided for most materials science specimens under well-defined diffraction conditions. Complex equations or methods are not required to account for nonlinear image formation. For a transmission electron microscope without corrector, the object wave function can be as usual obtained from the image wave function by taking into account the wave transfer function. A reliable reconstruction depends on the signal-to-noise ratio. Using silicon as a test object it was shown that a proper reconstruction is possible for electron doses as low as 250 electrons/pixel. The resolution of the reconstructed wave function corresponds to the resolution of the images, which is only marginally reduced during the reconstruction process.

\begin{acknowledgments}
The authors would like to thank B. Barton for stimulating discussions. The project is funded by the German Research Foundation (Deutsche Forschungsgemeinschaft) under Ge 841/16 and Sch 424/11.
\end{acknowledgments}

\bibliography{diss_main.bib}

\end{document}